\documentclass[prl,superscriptaddress,twocolumn,showpacs,%
amsmath,amssymb]{revtex4}
\usepackage{units}
\usepackage{graphicx}

\begin{document}

\title{Fine Structure of the Radial Breathing Mode in Double-Wall
Carbon Nanotubes}

\author{R. Pfeiffer}
\author{F. Simon}
\author{H. Kuzmany}
\affiliation{Universit\"at Wien, Institut f\"ur Materialphysik,
Strudlhofgasse 4, 1090 Wien, Austria}
\author{V. N. Popov}
\affiliation{Faculty of Physics, University of Sofia, Sofia, Bulgaria}

\date{\today}

\begin{abstract}
  The analysis of the Raman scattering cross section of the radial
  breathing modes of double-wall carbon nanotubes allowed to determine
  the optical transitions of the inner tubes. The Raman lines are
  found to cluster into species with similar resonance behavior. The
  lowest components of the clusters correspond well to SDS wrapped
  HiPco tubes. Each cluster represents one particular inner tube
  inside different outer tubes and each member of the clusters
  represents one well defined pair of inner and outer tubes.  The
  number of components in one cluster increases with decreasing of the
  inner tube diameter and can be as high as $14$.
\end{abstract}


\maketitle

Double-wall carbon nanotubes (DWCNTs) are new and interesting
structures in the family of carbon nanophases, especially when they
are grown by annealing of so-called peapods, i.\,e.\@ single-wall
carbon nanotubes (SWCNTs) filled with fullerenes
\citep{Bandow:ChemPhysLett337:48:(2001)}. The outer tubes diameters of
such DWCNTs are of the same small size (around $\unit[1.4]{nm}$) as it
is typical for SWCNTs. This means, they provide the same spatial
resolution for sensors or the same field enhancement for electron
emitters as the latter but exhibit higher stiffness. The inner tubes,
with typical diameters of $\unit[0.7]{nm}$, have a rather high
curvature and are thus expected to exhibit significant deviations from
graphene with respect to their mechanical and electronic
properties. Amongst others, the electron-phonon coupling is expected
to be considerably increased as compared to graphene, which might
result in superconductivity or a Peierls transition.  Also, the
concentric layers in the DWCNTs represent curved graphite sheets with
an interesting van der Waals type interaction and DWCNTs are well
defined model materials for multi-wall CNTs.  Finally, by filling
SWCNTs with isotope labeled fullerenes it is possible to produce
DWCNTs with $^{13}$C enriched inner and normal $^{12}$C outer tubes
\citep{Simon:PhysRevLett95:017401:(2005)}.

The peapod-grown DWCNTs attracted additional interest as the inner
tubes are grown in an highly unperturbed environment and thus exhibit
unusual narrow lines for the radial breathing mode (RBM) Raman
response \citep{Pfeiffer:PhysRevLett90:225501:(2003)}. In this sense,
the inside of the outer tubes has been termed a ``nano
clean-room''. Moreover, the Raman signal from the RBMs of the inner
tubes exhibits strong resonance enhancement and the observed number of
lines is larger than the number of geometrically allowed species
\citep{Pfeiffer:EurPhysJB42:345:(2004)}.

Similarly to the outer tubes, the geometric and electronic properties
of the inner tubes are uniquely determined by their chiral vectors
$(m,n)$, along which the graphene layers are rolled up into tubes
\citep{Reich:2004}. Due to the quasi one-dimensionality of the tubes,
the electronic states are jammed into Van Hove singularities
(VHSs). Electronic transitions of semiconducting ($E_{ii}^{\text{S}}$)
and metallic ($E_{ii}^{\text{M}}$) tubes are only allowed between
corresponding VHSs in the valence and conduction bands. These
transition energies scatter around lines with slopes proportional to
the inverse tube diameter $1/D$. Refined calculations revealed
considerable curvature corrections to the transition energies,
especially for small diameter tubes
\citep{Popov:NewJPhys6:17:(2004),Popov:PhysRevB70:115407:(2004),%
Samsonidze:ApplPhysLett85:5703:(2004)}. Experimental information on
these effects has recently been obtained from the Raman and
luminescence measurements of HiPco (high pressure carbon monoxide
grown) tubes with a mean diameter of about $\unit[1]{nm}$
\citep{Bachilo:Science298:2361:(2002),%
Fantini:PhysRevLett93:147406:(2004),Telg:PhysRevLett93:177401:(2004)}. The
pattern of the deviation from the linear relation between $E_{ii}$ and
$1/D$ can be categorized into tube families with respect to
$2m+n=\text{const}$. There are the metallic tubes with
$2m+n\equiv0\pmod3$ and two types of semiconducting tubes with
$2m+n\equiv1\pmod3$ (SI) and $2m+n\equiv2\pmod3$ (SII). For the
$E_{22}^{\text{S}}$ transition, the SI tubes deviate strongly from the
linear relation towards lower transition energies whereas the SII
tubes exhibit a smaller deviation towards higher transition
energies. In all cases zigzag-like tubes close to $(m,0)$ show the
largest deviation from the linear relationship.

The RBM frequencies also scale roughly as $1/D$ and have therefore
been used repeatedly for the structural analysis of SWCNTs,
particularly with respect to the diameter distribution. By analyzing
the Raman cross section of the RBMs it is possible to determine
chiralities and electronic transition energies of CNTs
\citep{Fantini:PhysRevLett93:147406:(2004),%
Telg:PhysRevLett93:177401:(2004)}.

\begin{figure*}
\includegraphics[width=0.75\linewidth]{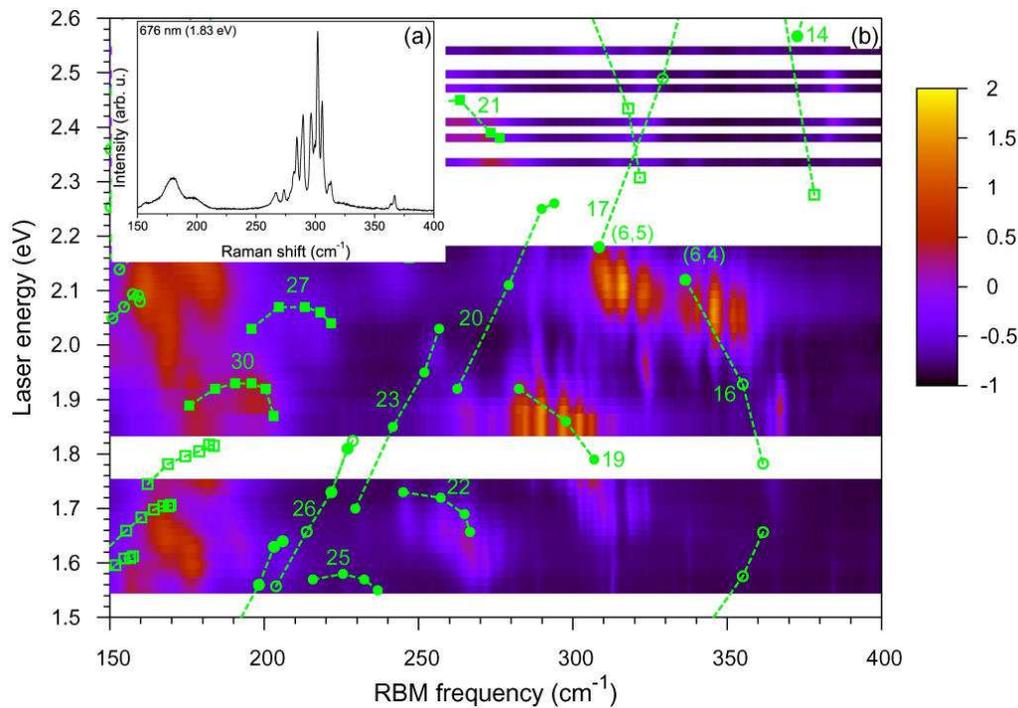}
\caption{(Color online) (a) Raman response of the RBM region of DWCNTs
  sample B. (b) Logarithm of the Raman cross-section of the tubes as a
  function of laser energy and RBM frequency. Filled symbols are
  measured RBM frequencies and transition energies of SDS wrapped
  HiPco tubes averaged from
  Refs.~\citep{Bachilo:Science298:2361:(2002),%
  Fantini:PhysRevLett93:147406:(2004),Telg:PhysRevLett93:177401:(2004)}
  (circles and squares denote semiconducting and metallic tubes,
  respectively). Open symbols are extended tight-binding calculations
  \citep{Popov:NewJPhys6:17:(2004)} corrected for many body effects
  after Ref.~\citep{Jorio:PhysRevB71:075401:(2005)}. Tube families are
  connected with dashed lines and assigned by family numbers
  $2m+n=\text{const}$.}
\label{Fig:KatPlot}
\end{figure*}

In this Letter, we present the results of resonance Raman measurements
in the RBM range of DWCNTs. These measurements unraveled the optical
transitions of the inner tubes which cannot be accessed by optical and
scanning probe experiments. Using a contour-plot where the Raman
cross-section is plotted over a 2D grid of RBM frequency and laser
excitation energy, we found a well expressed fine structure in the
resonance pattern. An analysis of this pattern showed a clustering of
the inner tubes RBMs near the HiPco transition energies and
frequencies. The clusters start around the HiPco frequencies and
extend up to $\unit[30]{cm^{-1}}$ to higher frequencies. They consist
of up to $14$ well expressed components. Each cluster represents one
particular inner tube inside different outer tubes and each member of
the clusters represents a well defined pair of one inner in one outer
tube. Within one cluster, the transition energies show a red-shift of
about $\unit[2]{meV/cm^{-1}}$. We explain the existence of such
``pair-spectra'' and the red-shift of the transition energies by the
interactions between the inner and outer tubes. This explanation is
supported by model calculations.

The DWCNTs studied here were obtained by annealing C$_{60}$ peapods at
$\unit[1250]{^\circ C}$ in a dynamic vacuum for $\unit[2]{h}$ as
described in Ref.~\citep{Bandow:ChemPhysLett337:48:(2001)}. Two
samples with mean outer diameters of $\unit[1.39]{nm}$ (sample A) and
$\unit[1.45]{nm}$ (sample B) were studied here. The Raman spectra for
the measurement of the resonance cross section were recorded in
backscattering geometry at ambient conditions using a Dilor XY triple
spectrometer operated in normal resolution and a LN$_2$ cooled CCD
detector. Selected spectra were measured in high resolution (HR) at
$\unit[90]{K}$. In this case, the resolution of the spectrometer was
$\unit[0.5]{cm^{-1}}$ for red laser excitation. The spectra were
obtained with different lasers such as an Ar/Kr, a Ti:Sapphire, and a
dye laser with Rhodamine 6G and Rhodamine 101. Excitation was between
$1.54$ and $\unit[2.54]{eV}$ ($488$ to $\unit[803]{nm}$) with a
spacing of about $\unit[15]{meV}$ in the red to yellow spectral
region. The frequencies were corrected using calibration lamps. In
order to determine Raman cross-sections and to correct for the
spectrometer/detector sensitivity, all spectra were normalized to the
well known cross-section for the Si $F_{1g}$ mode around
$\unit[520]{cm^{-1}}$.

Figure~\ref{Fig:KatPlot}(a) depicts a typical Raman spectrum of DWCNTs
in the frequency range of the RBM. The broad and structured line
pattern around $\unit[180]{cm^{-1}}$ and the set of strong and narrow
lines in the spectral range from $230$ to $\unit[400]{cm^{-1}}$
originate from the outer and from the inner tubes,
respectively. Figure~\ref{Fig:KatPlot}(b) shows the color coded
logarithm of the RBM Raman cross-section of DWCNTs sample B as a
function of laser energy and RBM frequency for normal spectrometer
resolution. The broad band between $150$ and $\unit[210]{cm^{-1}}$ is
the response from the outer tubes. Between about $230$ and
$\unit[400]{cm^{-1}}$ one can observe the RBM response of the inner
tubes. For this energy range, mainly the $E_{22}^{\text{S}}$ (circles)
and parts of the $E_{11}^{\text{M}}$ transitions (squares) can be
seen.

From the comparison with results for HiPco tubes one can easily
identify the families $22$, $19$, and $16$ in the SI branch of the
$E_{22}^{\text{S}}$ transition. Resonances from the SII tubes
(families $23$, $20$, and $17$) are in general weaker but can still be
observed, especially the $(6,5)$ tube in family $17$. The metallic
transitions are also close to the HiPco results.

The RBMs of the inner tubes clearly show a clustering of components
close to the HiPco transitions. The low frequency end of the clusters
coincides well with the results from the HiPco tubes except for a
downshift in energy of about $\unit[50]{meV}$. For the $(6,4)$ inner
tube in family $16$ the total width of the cluster is about
$\unit[30]{cm^{-1}}$. The width of the clustered lines clearly
decreases with increasing inner tube diameter. For the largest
diameter inner tubes (family $22$) almost no clustering is observed
and results are in very good agreement with the response from the
HiPco tubes.

A number of very intensive Raman lines are located close to the
transitions of the isolated $(6,5)$ and $(6,4)$ tubes in families $17$
and $16$, respectively. The two tubes are well separated in frequency
and the clusters can be well assigned to those tubes. In the following
discussion we will therefore mainly refer to these tubes, although all
results hold correspondingly for the other inner tubes. Within one
cluster the peak resonances of the components are clearly shifted to
lower energies with increasing frequency. For the $(6,5)$ and $(6,4)$
cluster we get $\unit[-2.7(6)]{meV/cm^{-1}}$ and
$\unit[-0.9(4)]{meV/cm^{-1}}$, respectively. On the average, the shift
is about $\unit[-2]{meV/cm^{-1}}$.

Two strong resonances at $(\unit[322]{cm^{-1}},\unit[1.92]{eV})$ and
at $(\unit[368]{cm^{-1}},\unit[1.86]{eV})$ could not be assigned,
although they were observed in all samples studied. The lower energy
resonance may be related to the $(7,2)$ tube of family $16$. The
higher energy resonance could originate from $(6,5)$ with a non
symmetric resonance transition
\citep{Grueneis:ChemPhysLett387:301:(2004)}.

\begin{figure}
\includegraphics[width=\linewidth]{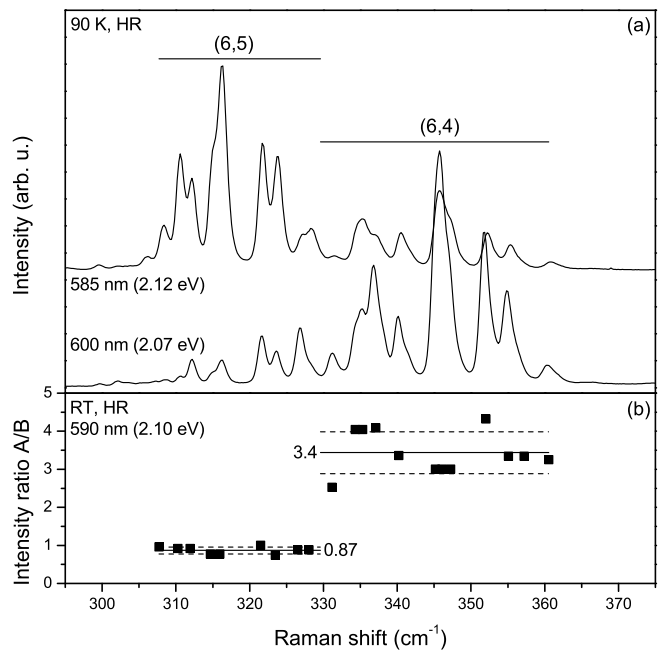}
\caption{(a) High resolution Raman response for the $(6,5)$ and
  $(6,4)$ inner tubes for sample A as excited with two different
  lasers. The horizontal lines cover the widths of the clusters. (b)
  Intensity ratio of the RBM components for the $(6,5)$ and $(6,4)$
  tubes of samples A and B with small and large mean outer tubes
  diameters, respectively. The dashed lines indicate the standard
  deviation.}
\label{Fig:IT6564}
\end{figure}

In order to demonstrate the clustering effect more clearly and to get
some information on the number of components in the clusters, we
measured the spectral range of the RBM for the $(6,5)$ and for the
$(6,4)$ tubes in high resolution at low temperature (see
Fig.~\ref{Fig:IT6564}(a)).
While for the lower energy laser all components of the $(6,4)$ species
are in resonance the components of the $(6,5)$ tubes remain weak. For
the higher energy laser the result is just opposite. From a fit to the
observed structures a total number of $9$ and $14$ components were
found for the $(6,5)$ and the $(6,4)$ clusters, respectively.

To check the stability of the clustered components in the spectra we
compared the patterns for two samples with different outer tube
diameters. As Fig.~\ref{Fig:IT6564}(b) shows, the $(6,4)$ cluster is
about a factor $3.4$ stronger in intensity in sample A
($\unit[1.39]{nm}$) than in sample B ($\unit[1.45]{nm}$). The
intensity of the $(6,5)$ components in sample A is only about
$\unit[87]{\%}$ of that in sample B. This is as expected, since for
the smaller diameter outer tubes proportionally more smaller inner
tubes grow \citep{Simon:PhysRevB71:165439:(2005)}.

The crucial result that a rather large number of peaks in the RBM
range of the Raman spectra have the same resonance behavior is
surprising. It strongly suggests that all components of one cluster
originate from the same inner tube. This means, the inner tubes can be
accommodated in a rather large number of different outer tubes and the
wall to wall distance can be quite different from its optimum value
\citep{Hashimoto:PhysRevLett94:045504:(2005)}. Thus, the clustered
lines represent pair spectra of one inner tube inside several well
defined outer tubes.

\begin{figure}
\includegraphics[width=\linewidth]{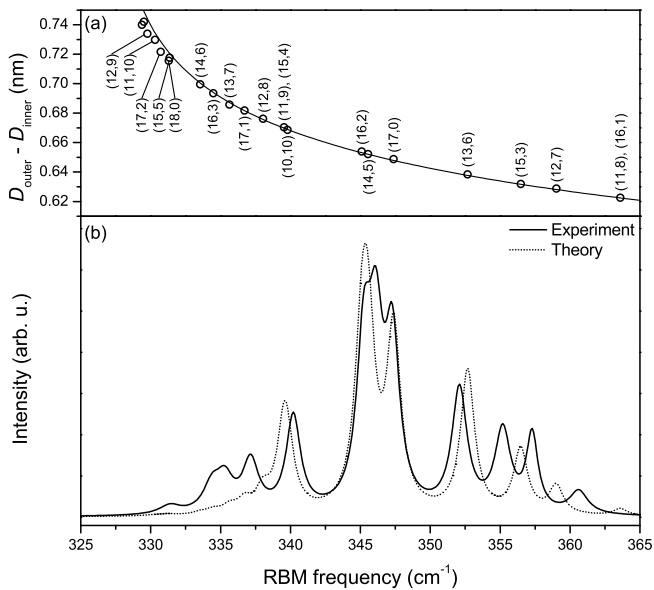}
\caption{(a) RBM frequency of a $(6,4)$ inner tube as a function of
  the diameter difference to various outer tubes (indicated by their
  chiral vectors). (b) Raman line pattern for the RBM of the $(6,4)$
  inner tube as obtained from experiment (solid line) and from theory
  (dotted line) as described in the text.}
\label{Fig:verteilung}
\end{figure}

Figure~\ref{Fig:verteilung}(a) depicts the calculated upshift of the
RBM frequency of a $(6,4)$ inner tube as a function of the diameter
difference to various outer tubes. In the calculations we considered
the tubes as elastic continuum cylinders interacting with each other
with a Lennard-Jones (LJ) potential parameterized for graphite
\citep{Lu:PhysRevLett68:1551:(1992)}. The relaxed structure of the
DWCNTs was found by minimization of the total energy $E$ equal to the
sum of the elastic energies of both tubes and the inter tube
interaction energy $E_{\text{LJ}}$, $E=Y_1\pi[R_{\text{i}}(\Delta
R_{\text{i}}/R_{\text{i}})^2+ R_{\text{o}}(\Delta
R_{\text{o}}/R_{\text{o}})^2]+E_{\text{LJ}}$. Here $R_{\text{i}}$ and
$R_{\text{o}}$ are the inner and outer tube radii, $\Delta
R_{\text{i}}$ and $\Delta R_{\text{o}}$ are their changes, $Y_1=Yd$,
$Y=\unit[1.06]{TPa}$ is the graphene in-plane Young's modulus, and
$d=\unit[0.335]{nm}$ is the tube wall thickness. Compared with the
continuum model results in
Ref.~\citep{Pfeiffer:EurPhysJB42:345:(2004)}, the frequency upshift
for small diameter differences became slightly smaller.

In Fig.~\ref{Fig:verteilung}(b) the solid line is the experimental
spectrum of sample A obtained in HR at $\unit[90]{K}$ with each
component normalized with the scattering cross section. The spectrum
thus shows the population distribution of the inner/outer tube
pairs. The calculated spectrum (dotted line in
Fig.~\ref{Fig:verteilung}(b)) was obtained by assigning a Lorentzian
to each calculated frequency whose intensity was for simplicity
assumed to be determined by a Gaussian distribution. The calculated
frequencies were scaled with a factor $0.98$. To get the best
agreement between theory and experiment, the parameters of the
Gaussian were $\omega_0=\unit[348.9]{cm^{-1}}$ and
$\sigma=\unit[5.2]{cm^{-1}}$. For the intermediate frequencies,
experiment and theory are in good agreement. For lower and higher
frequencies, the components deviate from the experiment. The maximum
of the simulated spectrum is found for a diameter difference of about
$\unit[0.66]{nm}$, which is smaller than the experimental value of
$\unit[0.72]{nm}$ reported in
Ref.~\citep{Abe:PhysRevB68:041405(R):(2003)}. This shows, that the LJ
potential must be reparameterized for the curved sp$^2$ networks of
DWCNTs.

The frequency upshift and the transition energy downshift within one
cluster can be understood from an increasing interaction between the
two shells for decreasing diameter difference. This interaction acts
on the inner shell like a radial pressure. The pressure dependence of
$E_{22}$ for the $(6,4)$ and the $(6,5)$ inner tubes was calculated in
the extended tight-binding framework
\citep{Popov:NewJPhys6:17:(2004)}. The corresponding values are
$\unit[-7.8]{meV/GPa}$ and $\unit[-2.5]{meV/GPa}$. After
Ref.~\citep{Venkateswaran:PhysRevB68:241406(R):(2003)} the frequencies
shift with about $\unit[1.1]{cm^{-1}/GPa}$. Thus one gets
$\unit[-7.1]{meV/cm^{-1}}$ and $\unit[-2.3]{meV/cm^{-1}}$,
respectively. The signs and the order of magnitude compare well with
the experiment. Additionally, from the measured pressure induced shift
of the $E_{22}^{\text{S}}$ transition
\citep{Wu:PhysRevLett93:017404:(2004)} for a $(6,5)$ tube of
$\unit[-4]{meV/GPa}$ one gets $\unit[-3.6]{meV/cm^{-1}}$, which is
very close the experimental value of $\unit[-2.7]{meV/cm^{-1}}$.

In summary, we report on the optical transitions of very high curvature SWCNTs
accommodated inside host outer tubes as derived from their resonance Raman
cross section. The observed Raman lines are clustered and represent pair
spectra between one inner tube and several well defined outer tubes. The
number of components in the cluster increases with decreasing tube diameter.
The leading edges of the clusters correspond to the HiPco tubes with respect
to frequency and transition energy except for a red-shift of $\unit[50]{meV}$.
Additional red-shifts of about $\unit[2]{meV/cm^{-1}}$ within a cluster are
understood from a radial pressure generated by different outer tubes. These
results provide now a unified picture for the high curvature nanotubes from
the HiPco process and from the DWCNTs species.

\begin{acknowledgments}
  Valuable discussions with A.\@ Jorio, A.\@ Rubio and L.\@ Wirtz are
  gratefully acknowledged. Work supported by FWF project P17345,
  Marie-Curie projects MEIF-CT-2003-501099 and MEIF-CT-2003-501080,
  and by NATO CLG 980422.
\end{acknowledgments}

\bibliography{cond-mat}

\end{document}